\documentclass[11pt]{article}
\usepackage{moriond,epsfig}

\def\Journal#1#2#3#4{{#1} {\bf #2}, #3 (#4)}

\def\be{\begin{equation}}
\def\ee{\end{equation}}
\def\bea{\begin{eqnarray}}
\def\eea{\end{eqnarray}}

\begin{document}
\vspace*{4cm}
\title{Preparation to the CMB Planck analysis : contamination due to the polarized galactic emission}

\author{L.~Fauvet, J.-F. Mac\'ias-P\'erez}

\address{LPSC, Universit\'e Joseph Fourier Grenoble 1, CNRS/IN2P3, Institut National Polytechnique de Grenoble, 53 avenue des Martyrs,\\ 38026 Grenoble cedex, France}

\maketitle\abstracts{
The Planck satellite experiment, which was launched the 14th of may 2009, will
 give an accurate measurement of the anisotropies of the Cosmic
Microwave Background (CMB) in temperature and polarization. This
measurement is polluted by the presence of diffuse galactic polarized
foreground emissions. In order to obtain the level
of accuracy required for the Planck mission it is necessary to deal
with these foregrounds. In order to do this, have develloped and
implemented coherent 3D models of the two main galactic polarized
emissions : the synchrotron and thermal dust emissions. We have
optimized these models by comparing them to preexisting data : the
K-band of the WMAP data, the ARCHEOPS data at 353 GHz and the 408 MHz
all-sky continuum survey. By extrapolation of these models at the
frequencies where the CMB is dominant, we are able to estimate the
contamination to the CMB Planck signal due to these polarized galactic
emissions.}

\section{Introduction}

The PLANCK satellite~\cite{bluebook} which is currently in flight and acquiring
data, should give the most accurate measurement of the anisotropies of the CMB
in temperature and polarization with a sensitivity of $\Delta T/T ~ 2\,\times
10^{-6} $ and an angular resolution of 5 arcmin.  Thanks to its range
of frequencies between 30 and 857 GHz it will give a great amount of
information about galactic and extra-galactic emissions. In order to
obtain the level
of accuracy required for the Planck mission it is necessary to deal
with these foregrounds and
the residual contamination due to these foreground emissions. While, for
the full sky, these emissions have the same order of magnitude than the CMB in
temperature, they dominate by a factor 10 in polarization~\cite{bluebook}. The
principal polarized Galactic microwave emissions come from 2 effects: thermal
dust emission and synchrotron emission. The synchrotron is well constrained by
the 408~MHz all-sky continuum survey~\cite{haslam}, by Leiden [Leiden/DRAO]
between~408 MHz and 1.4~GHz~\cite{wolleben}, by Parkes at
2.4~GHz~\cite{duncan1999}, by the MGLS {\it Medium Galactic Latitude Survey}
at 1.4~GHz~\cite{uyaniker} and by the satellite WMAP {\it Wilkinson Microwave
  Anisotropies Probe} (see e.g.~\cite{hinshaw}). The synchrotron emission is
due to ultrarelativistic electrons spiraling in the large-scale galactic
magnetic field. The thermal dust emission which has already been constrained
by IRAS~\cite{schlegel} and COBE-FIRAS~\cite{boulanger} is due to dust grains
heated by the the interstellar radiation field. Those grains emit a polarized submillimetric
thermal radiation~\cite{boulanger} as observed by e.g. ARCHEOPS~\cite{benoit}. The
polarization of these two types of radiation is orthogonal to the field
lines. We developed models of those emissions using the 3D galactic distribution of the magnetic field and
the matter density. The models are constrained using pre-existing data
and used to estimate the residual contamination to the CMB Planck
signal due to these polarized galactic emissions.

\section{3D modelling of the Galaxy}
\label{sec:model}

\indent A linearly polarized emission~\cite{kosowsky} at a given frequency
$\nu$ in GHz, can be described by the
Stokes parameters I, Q and U. For the polarized foreground emissions
integrated along the line of sight we obtain, for synchrotron:
\begin{eqnarray}
\label{eq:map_sync}
\centering
 I_s &=& I_{\mathrm{Has}} \left(\frac{\nu}{0.408}\right)^{\beta_s},\\
Q_s &=& I_{\mathrm{Has}}
\left(\frac{\nu}{0.408}\right)^{\beta_s}\frac{\int \cos(2\gamma)p_s n_e\left(B_l^2 + B_t^2 \right)}{\int n_e\left(B_l^2 + B_t^2 \right)} ,\\
U_s &=& I_{\mathrm{Has}}
\left(\frac{\nu}{353}\right)^{\beta_s}\frac{\int \sin(2\gamma)p_s n_e\left(B_l^2 + B_t^2 \right)}{\int n_e\left(B_l^2 + B_t^2 \right)},
\end{eqnarray}

\noindent where $B_n$, $B_l$ and  $B_t$ are the magnetic field components
along, longitudinal and transverse to the ligne of sight. The
polarization fraction $p_s$ is set to 75\%. $I_{Has}$ is the template
map~\cite{haslam}. The maps are extrapolated at all the Planck
frequencies using the spectral index $\beta_s$ which is a free parameter of the model.

For the thermal dust emission the Stokes parameters are given by: \\
\begin{eqnarray}
\centering
 I_d &=& I_{sfd} \left( \frac{\nu}{353} \right)^{\beta_d},\\
Q_d &=& I_{sfd} \left( \frac{\nu}{353}\right)^{\beta_d} \int n_d\frac{\cos(2 \gamma) \sin^2(\alpha) f_{\mathrm{norm}}p_d}{n_d} ,\\
U_d &=& I_{sfd} \left(\frac{\nu}{353}\right)^{\beta_d}  \int n_d \frac{\sin(2 \gamma) \sin^2(\alpha) f_{\mathrm{norm}}p_d}{ \int n_d} ,
\end{eqnarray}

\noindent where the polarization fraction $p_d$ is set to 10 \%, $\beta_d$ is
the spectral index (set to 2.0) and $f_{norm}$ is an empirical factor, fitted to
the ARCHEOPS data. The
$I_{sfd}$ map is the model 8 of~Schlegel {\it et al.}\cite{schlegel}.\\

\indent The models are based on an exponential distribution of relativistic
electrons on the Galactic disk, following~\cite{drimmel}, where the radial
scale $h_r$ is a free parameter. The distribution of dust grains $n_d$ is also
exponential~\cite{benoit}. The Galactic magnetic field is composed a regular and a turbulent components. The regular component is
based on the WMAP team model~\cite{page} which is close to a logarithmic
spiral to reproduce the shape of the spiral arms~\cite{han2006}. The pitch
angle $p$ between two spiral arms is a free parameter of the model. The
turbulent component is described by a Kolmogorov's law~\cite{han2006} 
spectrum with a relative amplitude $A_{turb}$.

\section{Comparison to data}
\label{sec:test}

\begin{figure}[htbf]
  \centering
  \psfig{figure=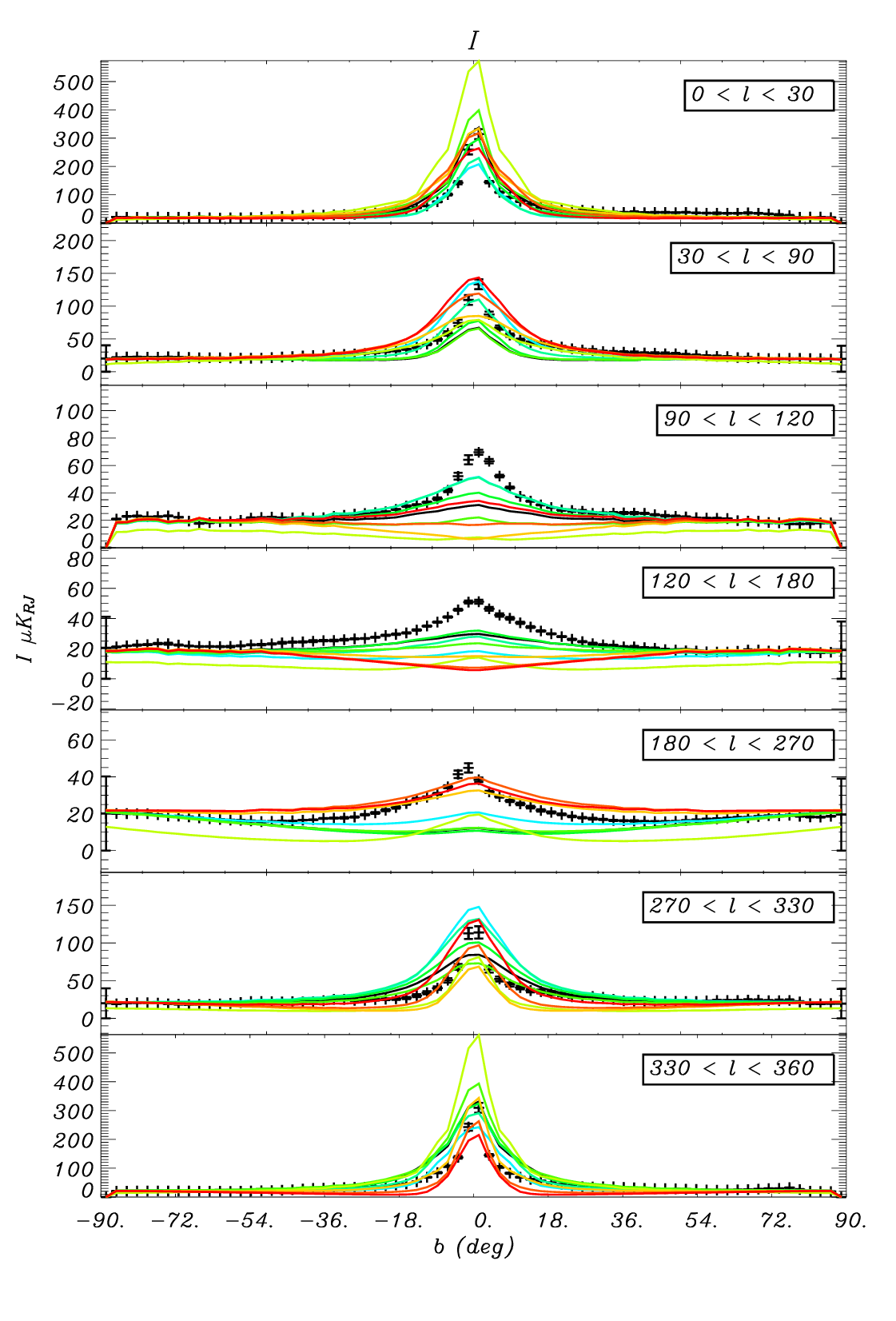,height=7cm,width=5cm}\caption{Galactic
    profiles in temperature at 408~MHz Haslam data in black and our
    synchrotron emission model for various values of the pitch angle $p$ {\it
      (form green to red)}.\label{fig:gal_has}}
\end{figure}

\indent We computed Galactic profiles in temperature and polarization for
various bands in longitude and latitude and various values of the free
parameters. In order to optimize these 3D models we compare them to Galactic
profiles computed with preexisting data using a $\chi^2$ test. For the
synchrotron emission in temperature, we use the 408~MHz all-sky continuum
survey~\cite{haslam} as shown on Figure~\ref{fig:gal_has}. In polarization we
use the K-band WMAP 5 years data. The thermal dust emission model is
optimized using the polarized ARCHEOPS data~\cite{benoit} at 353~GHz.

\indent The best fit parameters for the 3D model in polarization are given in
Table~\ref{tab:param}. The results are consistent for the three sets of
data. In particular we obtain compatible results for the synchrotron and
thermal dust emission models. $A_{turb}$ and $h_r$ are poorly constrained as
was already the case in Sun {\it et al}~\cite{sun}. The best fit value of the
pitch angle $p$ is compatible with results obtained by other
studies~\cite{sun,page}.  The best fit value for the spectral index of the
synchrotron emission is lower than the value found by ~\cite{sun,page}, but
this is probably due to the choice of normalisation for the regular component
of the magnetic field. With these models we reproduce the global structure of
the data (see for instance the Figure~\ref{fig:gal_has}) apart from the
Galactic Center.

\begin{table*}[h]
\begin{center}
\caption{Best fit parameters for synchrotron and thermal dust emission models.\label{tab:param}}
\vspace{0.3cm}

\begin{tabular}{|c|c|c|c|c|c|} \hline
$$  & $ p (deg)$& $A_{turb} $  & $h_r$  &    $\beta_s$  & $\chi^2_{min}$  \\\hline
$WMAP$ & $ -30.0^{+40.0}_{-30.0}$ & $< 1.25$ (95.4 \% CL) & $ >1$ (95.4 \%
CL) &  $-3.4^{+0.1}_{-0.8}$ & $5.72$     \\\hline
$HASLAM$ & $ -10.0^{+70.0}_{-60.0}$   & $< 1.25$ (95.4 \% CL) &  $ 5.0^{+15.0}_{-2.0} $ & $\emptyset$ & $5.81$ \\\hline
$ARCHEOPS$ & $ -20^{+80}_{-50}$   & $ < 2.25 (95.4 \% CL)$ & $\emptyset$ & $\emptyset$ &  $ 1.98$          \\\hline

\end{tabular}
\end{center}
\end{table*}

\section{Conclusions}

\begin{figure}[htbf]
\centering
\psfig{figure=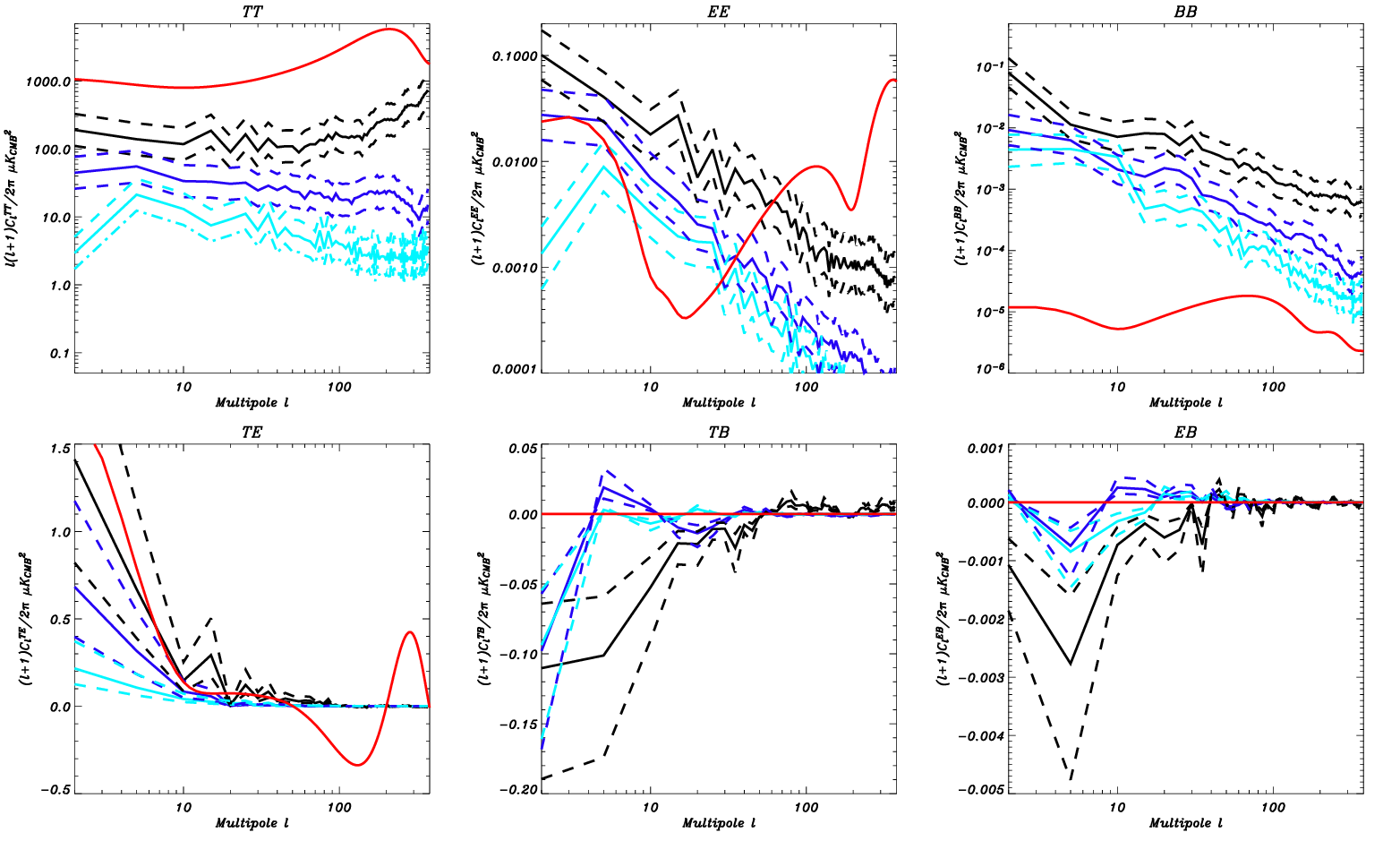,height=7cm,width=11cm}
\caption{Clockwise from top left : power spectra
  $C^{TT}_l$,$C^{EE}_l$,$C^{BB}_l$,$C^{TE}_l$,$C^{TB}_l$,$C^{EB}_l$ at
  143 GHz applying a galactic cut of $|b|<15^{\circ}$ (\emph{black}),
  $|b|<30^{\circ}$ (\emph{blue}) and  $|b|<40^{\circ}$ (\emph{cyan})   (see text for details).\label{fig:spect_cmb_for}}
\end{figure}

\indent From the above best fit parameters we estimated the contamination of
the CMB PLANCK data by the polarized galactic emissions. We compared power
spectra computed with simulations of the CMB PLANCK data \footnote{We used
  cosmological parameters for the $\Lambda$CDM--like model proposed in
  ~\cite{komatsu} with a tensor to scalar ratio of
  0.03.}. Figure~\ref{fig:spect_cmb_for} shows the temperature and
polarization power spectra at 143~GHz for the CMB simulation (\emph{red}) and
the Galactic foreground emissions, obtained by applying a Galactic cut $|b|<15^{\circ}$ (\emph{black}). The foreground contamination seems to be weak but for the
B modes an accurate foreground subtraction is extremely important concerning the
detection of the primordial gravitational waves. More details can be
found in~\cite{fauvet}.

\section*{References}
\begin{thebibliography}{99}

\bibitem{benoit} A. Beno\^it {\it et al}, \Journal{A\&A}{424}{512}{2004}
\bibitem{bluebook} The Planck Consortia, 2004, {\it Planck : The Scientific Program}
\bibitem{boulanger} F. Boulanger {\it et al}, \Journal{A\&A}{312}{256}{1996}
\bibitem{drimmel} R. Drimmel \& D.N. Spergel, \Journal{ApJ}{556}{181}{2001}.
\bibitem{duncan1999} A. Duncan et al, \Journal{A. \& A.}{350}{447}{1999}.
\bibitem{fauvet} L. Fauvet, J.F. Mac\'ias-P\'erez, F.X. D\'esert
  \emph{et al.}, {\bf astro-ph/1003.4450}.
\bibitem{han2006} J. L. Han {\it et al}, \Journal{A\&A}{642}{868}{2006}.
\bibitem{haslam} C.G.T Haslam et al{\it et al}, \Journal{A\&AS}{47}{1}{1982}.
\bibitem{hinshaw} G. Hinshaw {\it et al}, \Journal{ApJS}{170}{288}{2007}.
\bibitem{komatsu} E. Komatsu {\it et al}, \Journal{ApJS}{180}{306}{2009}
\bibitem{kosowsky} A. Kosowsky, \Journal{Ann. Phys.}{246}{49}{1996}.
\bibitem{dusta} M. -A. Miville-Desch\^enes {\it et al}, \Journal{A\&A accepted}{astro-ph/08023345}{2008}.
\bibitem{page} L. Page {\it et al},\Journal{ApJSS}{170}{335}{2007}.
\bibitem{schlegel} D. J. Schlegel, D. P. Finkbeiner \& M. Davies, \Journal{ApJ}{500}{525}{1998}.
\bibitem{sun} X.H. Sun {\it et al},\Journal{A \&A manuscrit}{astro-ph/0711.1572v1}{2008} .
\bibitem{uyaniker} B. Uyaniker {\it et al}, \Journal{A \& A.S.S. accepted}{astro-ph/9905023v1}{1999}.
\bibitem{wolleben} M. Wolleben {\it et al}, \Journal{A.\& A.}{448}{411}{2006}. 

\end{thebibliography}
\end{document}